\documentclass[%
 reprint,
 amsmath,amssymb,
 aps,
]{revtex4-2}

\usepackage{graphicx}
\usepackage{dcolumn}
\usepackage{bm}
\usepackage[normalem]{ulem}
\usepackage{multirow}
\usepackage[format=plain, justification=RaggedRight]{caption} 
\usepackage[htt]{hyphenat} 
\usepackage{subcaption}
\usepackage{subfig}
\usepackage{gensymb}
\usepackage{color}

\begin{document}

\preprint{APS/123-QED}

\title{Hindered stokesian settling of discs and rods}

\author{Yating Zhang}
\author{Narayanan Menon}%
\affiliation{%
 Department of Physics, University of Massachusetts, Amherst, Massachusetts 01003, USA\\
}%

\date{\today}


\begin{abstract}
We report measurements of the mean settling velocities for suspensions of discs and rods in the stokes regime for a number of particle aspect ratios. All these shapes display ''hindered settling'', namely, a decrease in settling speed as the solid volume fraction is increased. A comparison of our data to spheres reveals that discs and rods show less hindering than spheres at the same relative interparticle separation. The data for all six of our particle shapes may be scaled to collapse on that of spheres, with a scaling factor that depends only on the volume of the particle relative to a sphere. Despite the orientational degrees of freedom available with nonspherical particles, it thus appears that the dominant contribution to the hindered settling emerges from terms that are simply proportional to the volume of the sedimenting particles, enabling prediction of hindered settling of other nonpolar axisymmetric shapes.
\end{abstract}

\maketitle




Sedimentation, the gravitational settling of solid particles in a liquid, presents a complex challenge in classical many-body physics, particularly in the highly-viscous, overdamped Stokes regime where particle velocity fields decay very slowly with distance $r$ from the particle, as $1/r$. Many puzzles have emerged in studies of the sedimentation of suspensions of spheres \cite{chaikin2000thermodynamics,caflisch1985variance,nicolai1995particle,xue1992diffusion,mucha2004model}, especially on the scaling of fluctuations with system size. In this article, we address a basic question on the mean sedimentation rate in the collective settling of suspensions of \textit{nonspherical} particles, which are far more common in natural and technological sediments, such as clays, snowflakes, ice crystal, minerals, phytoplankton, reaction precipitates or paper fibres.

\begin{figure*}[!t]
\centering

\includegraphics[width=0.4\textwidth]{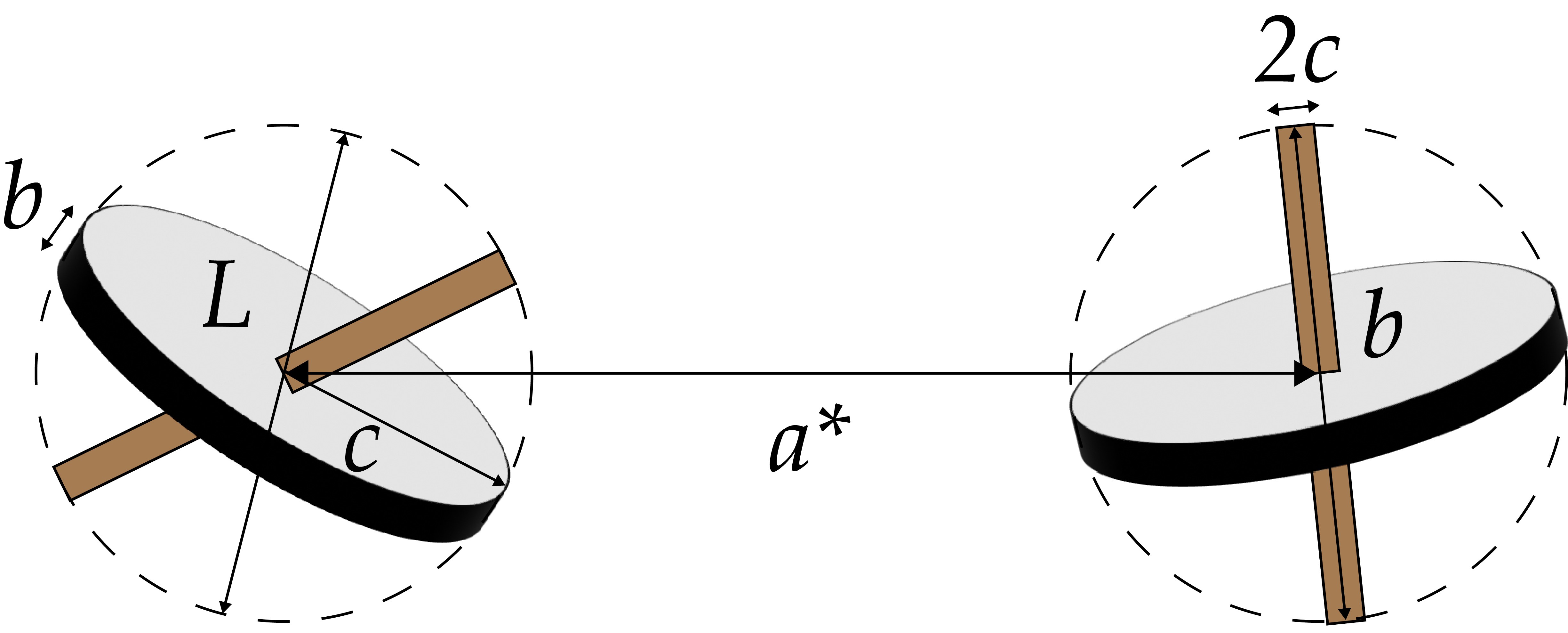}
\caption{\label{fig:Fig_1} Data for discs and rods are compared to those of spheres held at the same normalized interparticle separation $a^*=a/L$, where $a$ is the mean interparticle separation and $L$ is the largest particle dimension, i.e. diameter $2c$ of discs or the length $b$ of rods. Though the hex particle is not axisymmetric, its single-particle dynamics is the same as a disc; $L=2c$ is chosen to be the diameter of the circumcircle.}

\vspace{10pt} 


\captionof{table}{\label{tab:table1}%
Particle geometry, measured Stokes velocity, Reynolds number and P\'eclet number
}
\begin{ruledtabular}
\begin{tabular}{lccccc}
\textrm{Aspect ratio, shape}&
\textrm{c (mm)}&
\textrm{b (mm)}&
\textrm{$U_S$ (mm/s)}&
\textrm{Re}&
\textrm{Pe}\\
\colrule
A=0.15 hex & 0.27$\pm$0.01 & 0.08$\pm$0.001 &  0.04$\pm$0.001 & $6.33\times 10^{-5}$ & $1.07\times 10^{10}$ \\ 
A=0.07 hex & 0.65$\pm$0.02 & 0.09$\pm$0.003 & 0.13$\pm$0.003 & $4.96\times 10^{-4}$ & $1.36\times 10^{11}$\\ 
A=0.05 disc & 0.66$\pm$0.003 & 0.06$\pm$0.002 & 0.10$\pm$0.004 & $3.88\times 10^{-4}$ & $1.32\times 10^{11}$\\ 
A=19 rod & 0.15$\pm$0.0003 & 5.69$\pm$0.012 & 0.36$\pm$0.004 & $4.25\times 10^{-4}$ & $6.65\times 10^{12}$\\ 
A=10 rod & 0.26$\pm$0.0009 & 5.45$\pm$0.030 & 0.42$\pm$0.004 & $2.37\times 10^{-4}$ & $2.07\times 10^{13}$\\ 
A=3 rod & 0.49$\pm$0.0004 & 2.97$\pm$0.015 & 0.89$\pm$0.001 & $2.74\times 10^{-4}$ & $4.34\times 10^{13}$\\ 
\end{tabular}
\end{ruledtabular}

\end{figure*}
The question we address is how the settling velocity $U_S$ of an isolated particle of typical size $\ell$ in a fluid of dynamic viscosity $\eta$ and density $\rho$ is modified by the presence of a finite volume fraction $\phi$ of other particles. Our focus is on the Stokes regime where the Reynolds number $Re=\rho U_S \ell/\eta\ll1$. 
This effect is captured by the "hindered settling function" $H(\phi)$, a dimensionless ratio defined as $H=U/U_S$, where $U=U(\phi)$ is the mean sedimentation velocity of particles in suspension and $U_S$ is the stokes velocity of an isolated particle. As the name implies, in suspensions of spheres,  settling is hindered, with a monotonic decrease from $H(0)=1$ as volume fraction $\phi$ is increased \cite{davis1985sedimentation,ramaswamy2001issues}. 
Brzinski and Durian \cite{brzinski2018observation} made an extensive compilation of data for  sphere-sedimentation from previous studies. These studies include data for both small $Pe$ (brownian) and large $Pe$ (non-brownian), where the P\'eclet number $Pe=U_S\ell/D$ with diffusion coefficient $D$, quantifies the relative efficacy of convective to diffusive transport. These data are typically fit by the empirical Richardson-Zaki form $H(\phi)=(1-\phi)^n$ \cite{richardson1997sedimentation} with best-fit values of $n\approx 4.5$ for the large $Pe$, non-brownian regime that we consider in this article.  
There is no theory of $H(\phi)$ over the entire range of volume fractions; indeed, the only rigorous hydrodynamic calculation owes to  Batchelor \cite{batchelor1972sedimentation} who predicted that in the dilute limit, $\phi\rightarrow{0}$,  $H(\phi)=1-6.55\phi+\mathcal{O}(\phi^2)$.   



Shape leads to a qualitative difference in single particle dynamics and pair interactions: while a single sphere falls straight down without rotation, rods, discs, and ellipsoids drift horizontally and have vertical settling speeds that depend on their orientation relative to gravity \cite{taylor1969motion,batchelor1970slender,mackaplow1996study,mackaplow1998numerical,happel2012low}.
A pair of spheres falls faster than one sphere, remaining at fixed separation \cite{happel1960motion}, however pairs of discs settling at low $Re$ display either bound periodic orbits or scattering states \cite{chajwa2019kepler,jung2006periodic} due to the coupling between orientational and translational degrees of freedom. Likewise, a sedimenting lattice of discs shows very different collective behaviour \cite{chajwa2020waves} than a lattice of spheres \cite{crowley1971viscosity}.

However, the central question of this article is whether the effects of orientational degrees of freedom lead to a significant difference in collective behaviour of a finite volume fraction -- as opposed to a finite number -- of shaped particles compared to sphere suspensions, where many more studies exist. A prominent example of previous work on rods comes from  Guazzelli and coworkers \cite{herzhaft1996experimental,herzhaft1999experimental,guazzelli2011fluctuations,metzger2005large,metzger2007experimental}, who identified two distinct regimes of sedimentation for fibers: at very low volume fractions they found sedimentation is accompanied by a complex pattern of instabilities, whereas at higher volume fraction, they observed steady hindered settling reminiscent of sphere-suspensions.
We are not aware of comparable studies for the geophysically relevant case of flat particles at finite volume fraction. 

In this Letter, we present new data on the hindered settling of flat shapes (discs and hexagons) and elongated shapes (rods) in the Stokes regime, as a function of both particle aspect ratio and solid volume fraction, while remaining in a semi-dilute regime. These shapes are nonpolar (i.e. head-tail symmetric) and effectively axisymmetric. Our principal observation is that the dominant contribution to the hindered function $H(\phi)$ emerges from terms proportional to the volume of a single particle, over the wide range of aspect ratios that we explored.

We report results for three different aspect ratios each of flat (discs, hexagons) and long shapes (rods). All particle dimensions are given in Table~\ref{tab:table1} with the radius $c$, thickness/length $b$, and aspect ratio $A=b/(2c)$ \cite{supp}. The error bars, measured by the standard error on a population of individual particles, show that the discs, hex, and especially the rods, are very monodisperse. The rods were stainless steel, and used after cleaning and demagnetization in a degaussing coil. The discs and hex particles were polyimide, and used as purchased. 







The particles are suspended in silicone oil of density $0.97 g/cm^3$ and various kinematic viscosities ($350, 5000$ or $10000$ $cSt$) chosen to produce  Reynolds number $Re$ of approximately $10^{-5}$ to $ 10^{-4}$ in all situations. We prepare samples in a known volume fraction $\phi$ and initialize them to be spatially homogeneous as described in the End Matter. 

 
\begin{figure*}
\centering
      \begin{subfigure}[b]{0.15\textwidth}
        \centering
        \includegraphics[height=5cm, width=\linewidth, keepaspectratio]
        {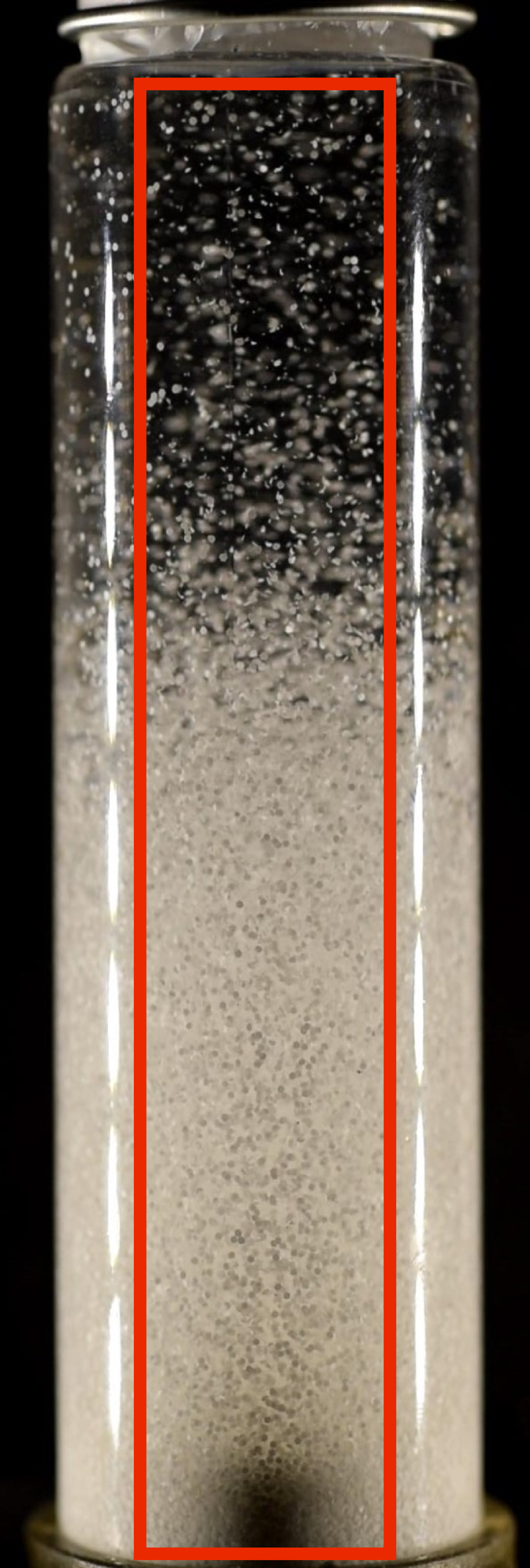}
        \caption{}
      \end{subfigure}
      \begin{subfigure}[b]{0.15\textwidth}
        \centering
        \includegraphics[height=5cm, width=\linewidth, keepaspectratio]
        {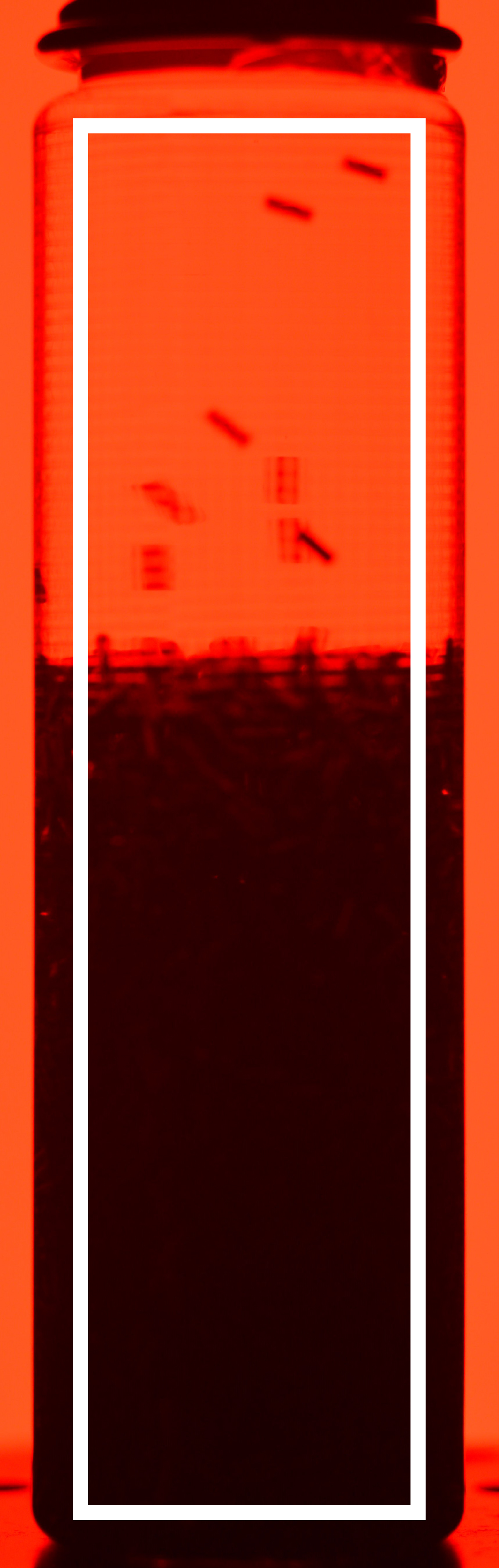}
        \caption{}
      \end{subfigure}
      \begin{subfigure}[b]{0.66\textwidth}
        \centering
        \includegraphics[height=5cm, width=\linewidth, keepaspectratio]
        {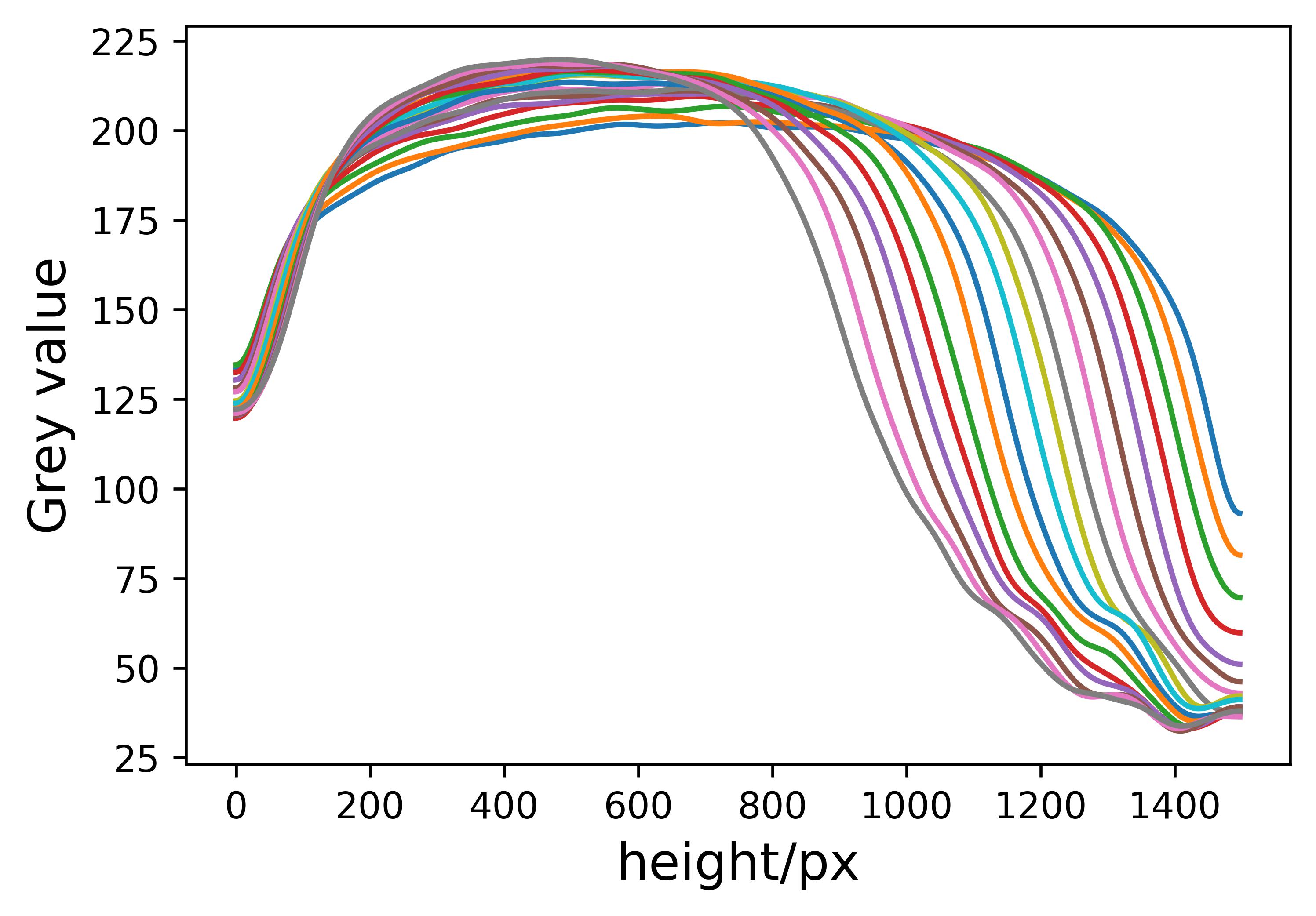}
        \caption{}
      \end{subfigure}
      
      \begin{subfigure}[b]{\textwidth}
        \centering
        \includegraphics[width=\linewidth]
        {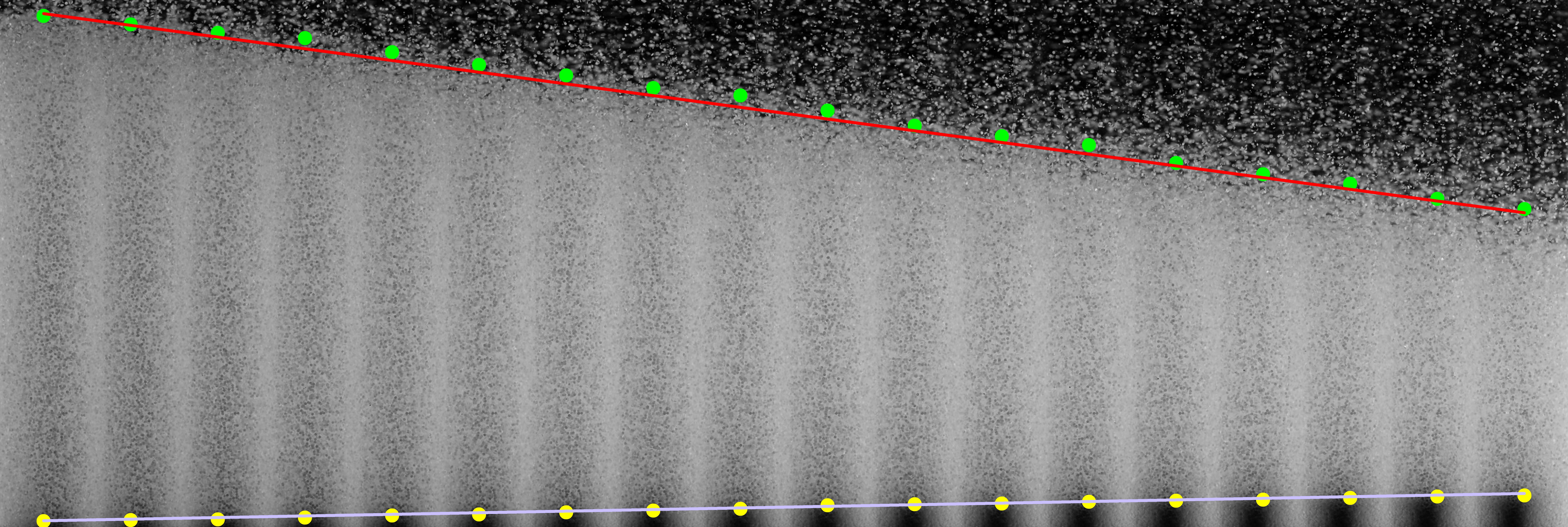}
        \caption{}
      \end{subfigure}
\caption{\label{fig:Fig_2}(a) Sample of $\phi=0.04, A=0.15$ hex particles imaged with scattered light. (b) Sample of $\phi=0.18, A=3$ rods imaged with transmitted light. The rectangles show the region selected for analysis. (c) Grey value of all selected images from (a), as a function of height at a series of times. The grey value is the intensity averaged over the horizontal direction. (d) Image sequence for the sample in (a).  The dots are the center points of interfaces found by analyzing the gradient of curves in (c). Green dots mark the interface detected between the suspension and the supernatant fluid. The yellow dots mark the interface between the suspension and the sediment solid. The red and purple lines are fits to the two sets of interface positions, whose slopes represent the sedimentation velocity $U$ and the velocity $U_p$ of the advancing solid precipitate, respectively.}
\end{figure*}
The specimens are quickly placed vertically with a precision of $\pm 0.2 \degree$. During sedimentation we took images with a Nikon DSLR camera of the entire cell at time intervals of typically $5 s$ until all solid particles settled at the bottom. For white particles (the hex and discs) we illuminated the sample from the side and imaged the scattered light, whereas for the opaque rods, we chose back-lighting to produce a high contrast, as shown in Fig.~\ref{fig:Fig_2}(a)\&(b). 

\begin{figure*}
\centering
    \begin{subfigure}[b]{0.8\textwidth}
        \centering
        \includegraphics[width=\linewidth]
        {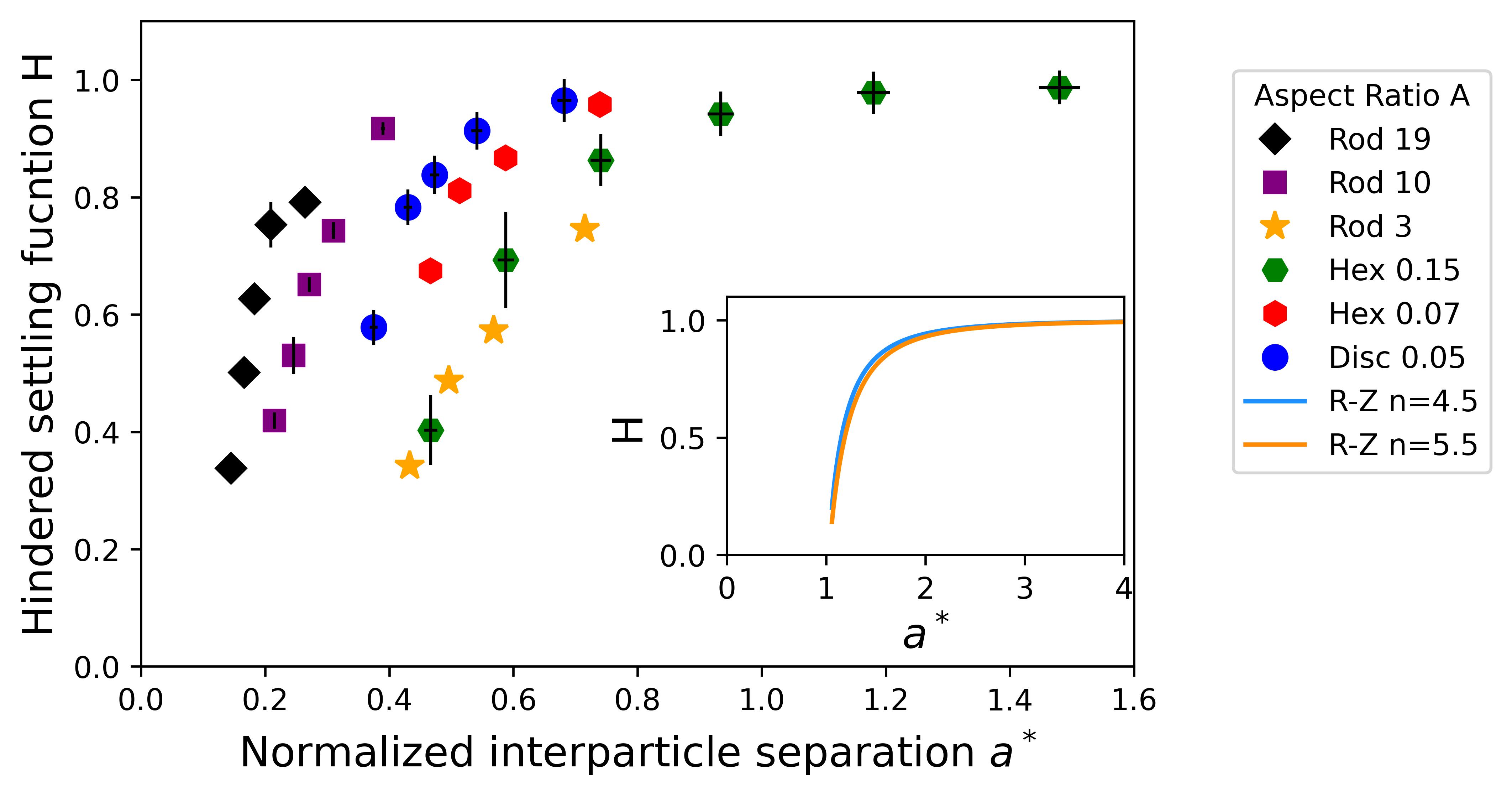}
        \caption{}
      \end{subfigure}
      
      \begin{subfigure}[b]{0.49\textwidth}
        \centering
        \includegraphics[height=5cm, width=\linewidth, keepaspectratio]
        {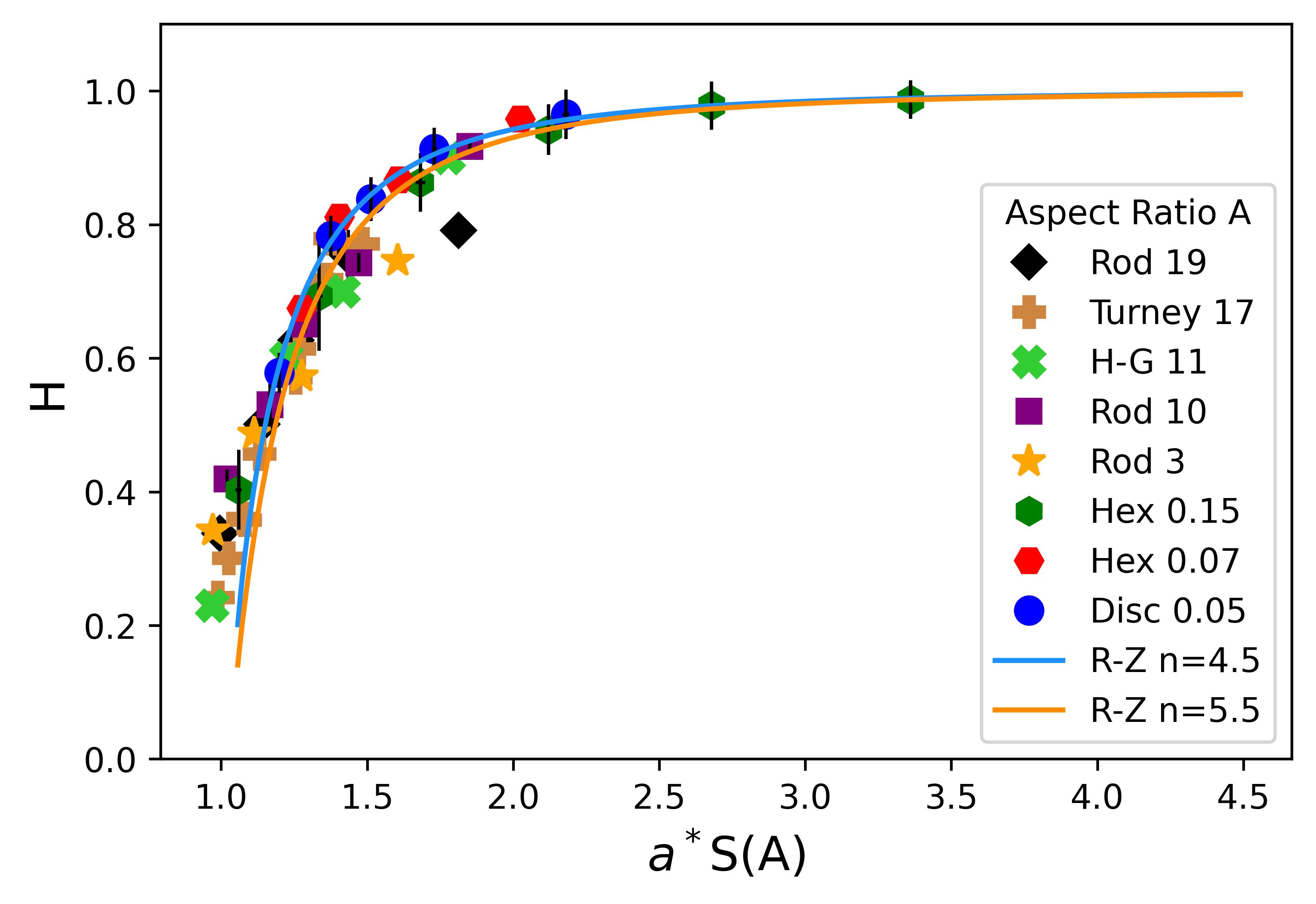}
        \caption{}
      \end{subfigure}
      \begin{subfigure}[b]{0.49\textwidth}
        \centering
        \includegraphics[height=5cm, width=\linewidth, keepaspectratio]
        {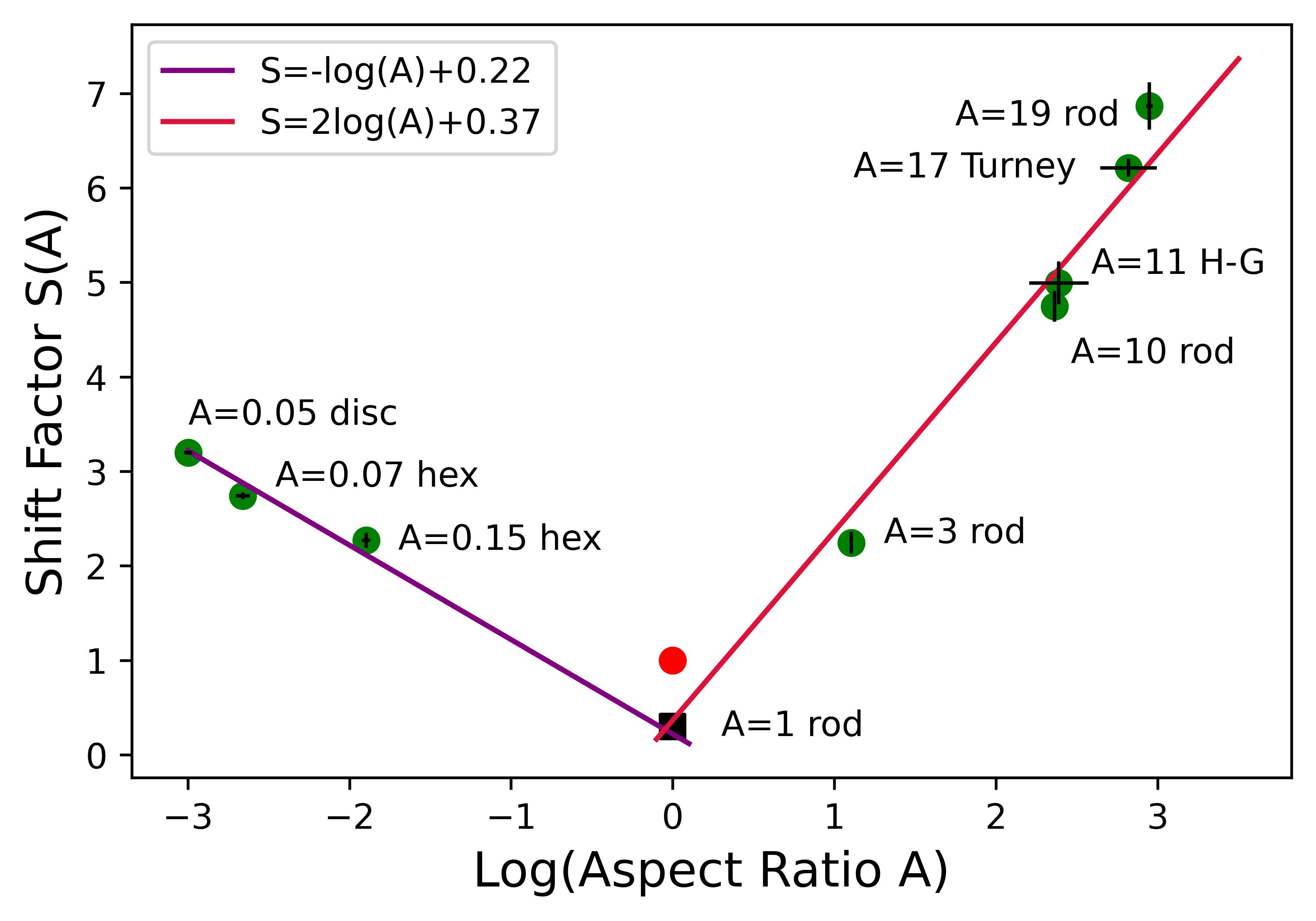}
        \caption{}
      \end{subfigure}
\caption{\label{fig:Fig_3}(a) Hindered settling function $H$ increases as normalized interparticle separation $a^*(\phi)$ increases for all particles we measured. Inset: The blue and orange curves are the Richardson-Zaki form $H=(1-\phi(a^*))^n$ with exponents n=4.5 \& 5.5 respectively \cite{richardson1997sedimentation}, which represent fits to sphere sedimentation.  (b) $H(a^*)$ data shifted by a factor $S(A)$ to collapse on the R-Z curve with $n=4.5$ based on the total least square method \cite{golub1973some,golub1980analysis}. We also include data from Herzhaft \& Guazzelli and Turney \textit{et al.} \cite{herzhaft1999experimental,turney1995hindered}. (c) Shift factor $S(A)$ versus the logarithm of the aspect ratio $A$ for both flat and long particles. The red dot for spheres is by definition at $S=1, A=1$. We also show a black square to indicate a cylinder with $A=1$, placed at the appropriate volume relative to the sphere. }
\end{figure*}

To obtain $U(\phi)$, the sedimentation velocity, we adapted a method based on Ref.~\cite{brzinski2018observation}. As shown in Fig.~\ref{fig:Fig_2}(a)\&(b), we select a central rectangular region,  and identify the position of the interface by taking gradients of the smoothed, horizontally-averaged intensity profile shown in Fig.~\ref{fig:Fig_2}(c). The outcome of this analysis is displayed in Fig.~\ref{fig:Fig_2}(d) where a time-series of images is stitched together. The upper green dots are the identified location of the interfaces between the suspension and the supernatant fluid.
The top interface is well-fit by a straight line (red line), implying that the sedimentation velocity $U(\phi)$ is constant over the entire settling process, and that the mean sedimentation is a steady-state process. This in turn implies that the sample was well-mixed initially and reproducibly prepared. 

For very low $\phi$, it is hard to detect the interface of between the suspension and the supernatant fluid when scattering or absorption of the illumination does not yield a sharply-defined interface.  Here we used Particle Image Velocimetry \cite{supp} to detect the velocity of individual particles and averaged over many particles to obtain the sedimentation velocity $U$.

All sedimentation data are tabulated in the Supplemental Material \cite{supp}. In the End Matter, we discuss finite-size effects on $U$, and measurements on the growth velocity of the suspension-precipitate interface (yellow symbols in Fig.~\ref{fig:Fig_2}(d)). 

To obtain the hindered function $H(\phi)$ from $U$, we normalize by the stokes velocity $U_S$. We determined the stokes velocity from measurements of individual particles dropped in a container much larger than the particle size. A complication is that $U_S$ for an anisotropic particle depends on orientation \cite{supp}. We measured $U_S$ for vertically oriented rods, but the flat particles were too small to fix an initial orientation, and the data averages over initial orientations. All the data are shown in Table~\ref{tab:table1}.


In Fig.~\ref{fig:Fig_3}(a) we show the principal experimental results of this article.  Rather than as a function of $\phi$, we plot the hindered settling function as a function of $a^*=a/L$, namely, the mean interparticle separation, $a$, normalized by the largest dimension $L$ of the particle, i.e., the diameter of hex and discs, and length of rods (see Fig.~\ref{fig:Fig_1}). The interparticle separation, $a$ is computed from the experimental volume fraction $\phi$ assuming a random spatial distribution of particle centres: $a^*=a/L=k \phi^{-1/3}$, where $k$ depends on aspect ratio $A$ \cite{hertz1909gegenseitigen,chandrasekhar1943stochastic,vroege1992phase,eppenga1984monte,veerman1992phase} as specified in the Supplemental Material \cite{supp}.

The data shown in Fig.~\ref{fig:Fig_3}(a) indicate a monotonically increasing trend for $H(a^*)$, showing that the effect of increasing volume fraction is indeed to hinder collective sedimentation, as with spheres. As a point of reference, we show in the inset curves for the Richardson-Zaki function with $n=4.5$ and $5.5$ which represent global fits \cite{brzinski2018observation} to data for spheres at $Pe>10^8$ and $Pe<10^8$, respectively. (For spheres, we use an expression for $a^*(\phi)$  valid to higher $\phi$ \cite{torquato1995nearest, supp}.) Deviations from $H=1$ appear at smaller normalized interparticle separation $a^*$ for discs and rods compared to spheres. Indeed, for shaped particles, hindering manifests for $a^*<1$, or $a<L$, in a semi-dilute regime where particles cannot rotate freely.
Said differently, with the same relative spatial distribution, non-spherical particles show much less hindering than spheres.

To more quantitatively compare between different aspect ratios, $A$, we shifted the data horizontally for each shape by a factor $S(A)$ to achieve the best possible collapse against spheres, by minimizing the total least squared deviation \cite{golub1973some,golub1980analysis} between the shifted data and the Richardson-Zaki function with $n=4.5$, as shown in Fig.~\ref{fig:Fig_3}(b). 
We also included data from Refs. \cite{herzhaft1999experimental,turney1995hindered} in Fig.~\ref{fig:Fig_3}(b), shifted in the same manner. To the extent that these eight data sets collapse within experimental error, the shift factor $S(A)$ captures the shape-dependence of the hindering effect of equally spaced particles with equal largest dimension.

To explore further this shape dependence, we plot in Fig.~\ref{fig:Fig_3}(c) the shift factor $S(A)$ against the logarithm of the aspect ratio $A$. $S(A)$ for $A<1$ (flat shapes) and $A>1$ (rods) are proportional to $A^{-1}$ and $A^{2}$ respectively, with prefactors of order unity. When scaled to the largest particle dimension $L$, these aspect ratio dependencies capture the relative volume of these shapes, %
with the volume of sphere, disc and rod expressed as $\pi/6, \pi A/4$ and $\pi /(4A^2)$ in units of $L^3$. 
Thus the scaling of the shift factor $S(A)$ as $A^{-1}$ for discs and $A^{2}$ for rods is exactly the ratio of the sphere volume to the volume of the shape.  We note that the point for spheres, set at $S(A)=1, A=1$, lies above the intersection of the rod and the disc scaling.  This is as expected, since the volume of the inscribed sphere is less than the volume of the cylinder with $A=1$ that should lie at the intersection point. 

The data thus point to the significance of particle volume. A clue to the dominant contributors to hindering are available in the exact calculation by Batchelor for spheres in the dilute limit ($\phi\rightarrow{0}$), where he found $H(\phi)=1-6.55\phi+\mathcal{O}(\phi^2)$ \cite{batchelor1972sedimentation}. This calculation identifies backflow as the dominant contribution to the hindering. When particles settle, there must be fluid displaced upwards equal to their volume. In addition, a settling particle also drags surrounding fluid down, which scales with the sphere volume. The upward flux of fluid is irrelevant for the stokes velocity of an isolated particle as the return flow is at infinity, but for a finite $\phi$, the upward backflow goes through the bulk of the suspension which amounts to a contribution of -5.5$\phi$ out of -6.55$\phi$. The remainder comes from hydrodynamic interactions that go beyond the mean flow: -1.55$\phi$ and +0.5$\phi$ come from the probability of close encounters between spheres and hydrodynamic interactions via the second derivative of the velocity field generated by other spheres \cite{batchelor1972sedimentation, chaikin2000thermodynamics}.  We are not aware of any other Batchelor-style calculations in the dilute limit for anisotropic particles, nor are there calculations beyond the dilute regime for the relative contributions of backflow versus interparticle hydrodynamic interactions. However, the combined data from spheres, discs and rods suggests that backflow dominates even at finite $\phi$.



Our data therefore suggest that the mean hindered settling of discs and rods in the stokes regime are to a first degree not more complicated than that of spheres, despite the orientational degrees of freedom that potentially add new physics to sedimentation dynamics. This observation enables prediction of the hydrodynamic contribution to hindering of any nonpolar axisymmetric shape, such as a paper fibre, a red blood cell, or a clay platelet. Subtler effects such as the sharpness of the sedimentation interface or fluctuations introduced by  orientational degrees of freedom remain to be explored.

Our conclusions are based on experiments in the semi-dilute regime where interparticle separation is of the order of the particle size. At lower volume fractions, rods have been shown to experience segregation into dense lanes of streamers and particle-depleted regions where the backflow can bypass particle-rich regions \cite{metzger2007experimental}. These instabilities appear to be system-size dependent and the boundary between the stable, hindered regime we studied and the unstable regime deserves further exploration. At higher volume fractions, translational and orientational degrees of freedom can become constrained due to both steric and hydrodynamic interactions and  more complicated physics can emerge.


We gratefully acknowledge financial support through NSF-DMR 2319881. We thank D.J. Durian, B. Metzger and V. Mathai for their help with the literature, and A. Rodrigues for her help with density measurements on discs.


\section{End Matter}

\textit{Determining volume fraction} To obtain a suspension of a known volume fraction $\phi$, we first measured the mass density of the solid particles using a pycnometer (EISCO Labs). We filled the pycnometer of $10\pm0.03 ml$ with distilled water to ascertain its precise volume. We then added a known mass of particles and a few drops of 1\% soap solution which suppressed bubbles from adhering to solid surfaces, then filled the container with distilled water. Using the measured density of the fluid, we calculated the mass density $\rho_s$ of the particles. The precise volume of the sedimentation cell $V$ is then measured, and a mass $m$ of particles is added. This leads to a sample of known volume fraction $\phi=m/(\rho_s V)$ with a precision of better than $\pm 2.7\%$, as described in the Supplemental Material \cite{supp}. The largest quantifiable source of error in $\phi$ comes from the measurement of the density $\rho_s$ of each of the types of particle. In addition, there may be additional small errors from particles adhering to surfaces, or from imperfect filling. Error in $\phi$ has a systematic effect on all volume fractions of a given type of particle. (Similarly, an error in $U_s$ is a systematic error in $H$ for a given particle type. Both these errors affect the quality of the scaling in Fig. 3, whereas random errors in $U$ do not.)

\textit{Sample preparation} We placed partially-filled sample containers with particles in silicone oil and a beaker of silicone oil in a vacuum chamber for more than 24 hours to remove bubbles. The sample containers were then nearly filled with silicone oil from the beaker. We applied vacuum again for several hours to remove the remaining bubbles and then overfilled the samples with silicone oil. Next, we sealed the bottles with Teflon tape and threaded lids. Once a sample was prepared, we put it in a home-built tube rotator for mixing overnight at low frequency (typically at a few r.p.m.) to generate a well-mixed homogeneous initial condition at the chosen volume fraction $\phi$.

\textit{Finite size effects}
Finite-size effects arising from the wall are expected to be smaller for the sedimentation velocity $U$ itself, due to screening of wall effects in the bulk. We confirm this expectation for a series of $\phi=0.04, A=0.15$ hex samples in containers with varying heights and widths as shown in the Supplemental Material \cite{supp}.

\textit{Precipitate volume fraction}
In addition to $U(\phi)$ for the receding top interface, for the flat particles imaged by scattered light, we also obtain the velocity $U_p$ with which the solid precipitate front grows, as shown by the the yellow dots and the purple line in Fig .~\ref{fig:Fig_2}(d). Mass conservation of the solid  implies $(U+U_p)\phi=U_p\phi_p$ which allows us to extract the volume fraction $\phi_p$ of the precipitate. A consistency check is provided by directly measuring $\phi_p$ from the height of the precipitate after sedimentation is complete; the agreement of the value of $\phi_p$ from these two methods \cite{supp} suggests that the measured interface velocity also tracks particle sedimentation dynamics.

\bibliography{main}

\end{document}